# Exchange coupled perpendicular media

D. Suess, J. Lee, J. Fidler
Institute of Solid State Physics, Vienna University of Technology, Austria

T. Schrefl
Department of Engineering Materials, The University of Sheffield, Sheffield, UK.

Abstract: The potential of exchange spring bilayers and graded media is reviewed. An analytical model for the optimization of graded media gives an optimal value of the magnetic polarization of $J_s = 0.8$ T. The optimum design allows for thermally stable grains with grain diameters in the order of 3.3 nm, which supports ultra high density up to 5 – 10 Tbit/inch². The switching field distribution is significantly reduced in bilayer media and graded media compared to single phase media. For the graded media the switching field distribution is reduced by about a factor of two. For bilayer media the minimum switching field distribution is obtained for soft layer anisotropies about 1/5 of the hard layer anisotropy. The influence of precessional switching on the reversal time and the reversal field is investigated in detail for magnetic bilayers. Exchange spring bilayers can be reversed with field pulses of 20 ps.

## 1. Introduction

The areal density in magnetic recording disks grows at the rate of about 40% a year [1]. Hard disk media that support ultra high densities require small grains in order to obtain high signal to noise ratios. The use of high coercive materials such as alloys in the $L1_0$ phase (e.g. FePt, FePt, CoPt, MnAl) allow for thermally stable grains at grain diameters in the order of 4nm [2]. However state of the art write heads cannot produce sufficiently large fields to reverse these extremely hard magnetic grains. Generally, the write field requirements to reverse the last grain of a written bit can be expressed as [3]

$$\mu_0 H_{sat} = \mu_0 H_{s,grain} + N_{eff} J_s - \mu_0 H_{ex} \qquad (1)$$

The first term, $\mu_0 H_{s,grain} = \mu_0 \alpha 2K / J_s$, takes into account for the intrinsic anisotropy and the shape anisotropy of one grain. It denotes the switching field of one grain of the media when no neighboring grains are present. The structure parameter $\alpha$ depends on the geometry of the grain and the angle between the write field and the easy axis. For tilted media for example, the external field is applied at an angle of 45° leading to $\alpha$ of about 1/2. $K$ and $J_s$ are the anisotropy constant and the magnetic polarization, respectively.

The second term takes into account for the stray field of the neighboring grains. If all grains except one grain in the centre of the bit are reversed the neighboring grains produce a considerable stray field that stabilizes the last grain. The stabilizing field can be expressed with an effective demagnetization factor $N_{eff}$, which is about 0.5 to 1.

The last term in Eq. (1) account for the exchange field that helps to reverse the last grain of a bit. The optimum value of the intergrain exchange field, which can be controlled by the intergranular phase is a complicated task, which requires the calculation of signal to noise ratios and calculations of the thermal stability. Gao and Bertram propose moderate values of the exchange field of about 5% of the anisotropy field[4]. In practice, it seems very probably that for perpendicular recording media with areal densities of about 500 Gbit/inch², much larger values of the intergrain exchange are used to achieve the required thermal stability. It seems that due to high head field gradients higher values of the intergranular exchange can be tolerated.

Recently composite media and exchange spring media were introduced to reduce the write field requirements. In exchange spring media a soft magnetic phase is perfectly exchange coupled to the hard magnetic grain. In composite media an intermediate layer lowers the exchange coupling between the hard and the soft phase. It was shown the required write field decreases by a factor of between two and four in composite media and exchange spring media, respectively [5,6]. The effect of different soft layer anisotropies on the switching field was studied in detail by Dobin et al [7]. Garcia-Sanchez et al. studied the reversal process of magnetic bilayers using a multiscale approach where the interface between the hard and soft layer is discretized atomistically. They found that strong coupling between the soft and the hard layer is required to obtain a maximum reduction of the switching field [8]. Experimental work on composite/exchange spring media can be found in Ref [9,10].

The paper is structured as following. In Section 2 basic concepts of magnetic bilayers are reviewed.

In Section 2.1 the influence of the soft layer anisotropy on the switching field is investigated for the highly damped case, where precessional switching can be neglected.

In Section 2.2 detailed studies are perfomed to investigate switching field distribution in bilayer media. The section deals with analytical investigations on the switching field distribution for infinite thick bilayers. Numerical studies are presented for more realistic bilayer with thicknesses of 20nm. Different cases between the correleation of anisotropy distributions between the soft layer and the hard layer are investigated.

The influence of fast field rise times and small damping is investigated in Section 2.3. A comparison with single phase media shows that bilayer media can be switched faster than single phase media.

Section 3 deals with graded media which is an extension of bilayer media to multilayer media. In Section 3.1 an analytical formula is developed that allows determining the optimal material parameters for high thermal stability and good writeability. The gain of graded media compared to bilayer media is discussed in Section 3.2.

In section 3.3 the origin of the reduced switching field distribution in graded media is discussed.

An optimistic outlook for a media design that allows areal densities up to 10Tbit/inch² is given in 3.4.

Open questions and challenges of bilayer media and graded media are discussed in section 4 and section 5.

## 2. Bilayer media

### *2.1. Switching field and thermal stability*

In exchange spring media the switching field of an extremely hard magnetic layer is reduced by a softer magnetic layer that is exchange coupled to the hard layer. The hard layer is columnar grown on the softer layer. In a first step a nucleation is formed in the soft layer as shown in Fig **1** . In the limit that the soft layer is strongly exchange coupled to the hard layer and provided that the thickness of the soft layer $t_s$ is larger than the domain wall width of the hard layer the nucleation field in the soft layer can be given as ,

$$H_n = 2K_{soft}/J_s + 2A\pi^2/(4t_s^2 J_s) \cdot \qquad (2)$$

where $K_{soft}$, $J_s$ and $A$ is the anisotropy, the spontaneous polarization and the intragranular exchange constant in the soft layer, respectively [11],[12]. The last term accounts for the exchange field from the hard layer that stabilizes the soft layer. The formation and expansion of the nucleation is reversible up to state (D) in Fig **1**.

In a second step the formed nucleation propagates to the soft/hard interface where it becomes pinned. With increasing external field strength the domain wall becomes compressed as shown in Fig **1**. The bilayer completely reverses when the external field is large enough to overcome the pinning field $H_p$. Therefore, the switching field of the entire grain is given by,

$H_{s,grain}$=max $(H_n, H_p)$.  (3)

The reduction of the switching field of one grain of an exchange spring media can be easily understood by an analytical formula that was derived by Hagendorn[13], in a more general case by Loxley et al.[14] and in the most general case by Kronmüller and Goll [15]. In the general case the pinning field of the bilayer structure is given by,

$$H_p = \frac{2K_{hard}}{J_{s,hard}} \frac{1-\varepsilon_K \varepsilon_A}{\left(1+\sqrt{\varepsilon_J \varepsilon_A}\right)^2}. \quad (4)$$

The ratio of the material parameters of the soft layer to the hard layer are given by the ratio of the anisotropy constants, $\varepsilon_K = K_{soft}/K_{hard}$, the ratio of the exchange constants, $\varepsilon_A = A_{soft}/A_{hard}$, and the ratio of the magnetic polarizations, $\varepsilon_J = J_{s,soft}/J_{s,hard}$. Eq. (4) shows that the pinning field can be reduced by increasing the magnetization in the soft layer and also by a large ratio of the exchange constant in the soft layer to the exchange in the hard layer.

The influence of the anisotropy constant in the soft magnetic part on the switching field of the columnar grain is shown in Fig **2**. The demagnetizing field which gives a small contribution to an additional uniaxial anisotropy is not taken into account. It is interesting to note that an inplane anisotropy in the soft layer is not beneficial in order to further reduce of the switching field. The smallest switching field is obtained for a soft layer anisotropy that is about 1/5 of the anisotropy of the hard layer anisotropy [13].

Besides the switching field the energy barrier between the state where the magnetization in the hard fraction is pointing up and the state with the hard layer magnetization pointing down is investigated using the nudged elastic band method [16]. For positive values of the anisotropy $K_1$ (here, we do not denote the top layer with $K_{soft}$ because we vary the anisotropy constant from large negative value to large positive values, which makes the expression soft layer misleading) the energy barrier is independent of $K_1$. The initial state is a state with homogenous magnetization which points up. The magnetic configuration at the saddle point shows a domain wall in the hard grain. The difference in energy between these two states is independent of $K_1$. The situation is different if an inplane anisotropy is assumed in the soft grain. The saddle point configuration does not contain a full 180° domain wall. Hence the energy barrier is smaller for negative values of $K_1$. Besides the energy barrier at zero field as investigated in Fig **2** the energy barrier under the influence of opposing external fields is important for at least three reasons. (i) $\Delta E(H)$ determines the decay of the magnetization of a bit in actual recording media, where bits are typically exposed to an effective external field, generated by neighboring bits. (ii) A knowledge of $\Delta E(H)$ is important to determine the zero field energy barrier from dynamic coercivity measurements. (iii) the reduction of the thermal stability of adjacent tracks due to the head field can be investigated with the knowledge of $\Delta E(H)$[17].

Detailed analysis of $\Delta E(H)$ can be for example found for single phase media in Ref[18,19,20], for bilayer media in Ref[21,22] and for graded media in Ref [23].

## *2.2. Switching field distribution*

In single phase media a variation of the anisotropy constant within each grain linearly changes the switching field.

$$\frac{\Delta H}{H} = \frac{\mu_0 H_{s,grain,SW}(K_1+\Delta K_1) - \mu_0 H_{s,grain,SW}(K_1)}{\mu_0 H_{s,grain,SW}(K_1)} = \frac{\Delta K_1}{K_1} \qquad (5)$$

**Infinite thickness of soft and hard layer**

The situation is different in sufficient thick bilayer media, where a full domain wall can be formed[24]. The switching field is determined by Eq.(3). In this case two situations have to be distinguished

1. $H_n > H_p$ : If the nucleation field is larger than the pinning field, the switching field of the particle is determined by the soft layer anisotropy only. During reversal the domain wall moves from the soft end into the hard end. Since in the considered case the pinning field is much smaller than the nucleation field any variations of the hard layer anisotropy that do not lead to a pinning field that is larger than the nucleation field, do not change the switching field of the entire grain. Only variations in the soft layer anisotropy (or variations of the thickness of the soft layer) will lead to a switching field distribution. The switching field depends linearly on the soft layer anisotropy.

2. $H_n < H_p$ : If the pinning field is much larger than the nucleation field, a reversed domain can be easily formed in the soft magnetic end. However, the domain wall becomes pinned at the soft/hard interface. If the external field exceeds the pinning field the entire particle can be reversed. The switching field depends linearly on the difference of the hard layer anisotropy and soft layer anisotropy.

## Finite thickness of soft and hard layer

If the soft layer is not sufficient thick forming a full domain wall, the influence of a variations of the anisotropy constant can not be easily estimated analytically. In this case numerical calculations are required. In the following we present results for one grain with a 10 nm thick soft layer and a 10 nm thick hard layer ( $A=10^{-11}$ J/m, $J_s=0.5$ T). The demagnetizing field is omitted. The demagnetizing field can be taken into account by an addition contribution to the uniaxial anisotropy.

Fig **3** shows the dependence of the switching field for one grain of the bilayer for different soft layer anisotropy constants ($K_{soft}$) as function of the hard layer anisotropy constant ($K_{hard}$). We again distinguish two cases:

1. $H_n > H_p$ : The transition where the nucleation field becomes larger than the pinning field can be approximated by comparing Eq. (2) and (4). For sufficient hard layers (e.g. $K_{soft} = 1$ MJ/m³ ) the transition occurs approximately if $K_{hard} < 5\, K_{soft}$ . The previously discussed insensitivity of the switching field on the hard layer anisotropy can be best seen for $K_{soft} = 1$ MJ/m³ in Fig **3**. However, this insensitivity does not already imply that we have decreased the switching field distribution. The switching field can strongly depends on the soft layer anisotropy.

2. $H_n < H_p$ : Approximately for $K_{hard} > 5\, K_{soft}$ the switching field increases linearly with the hard layer anisotropy, which is in agreement with the previous discussion.

## Simulations on ensembles of non-interacting grains

In order to show how variations of the anisotropy constants in the soft and hard layer influence the switching field, calculations on ensembles of non-interacting grains are performed. The switching field distribution of 400 grains was numerically calculated. The mean value of the hard layer anisotropy was assumed to be $K_{hard} = 1$ MJ/m³ . The mean value of the anisotropy constant in the soft layer was varied. The demagnetizing field is omitted. The exchange constant and the magnetic polarization was assumed to be the same in the soft layer and the hard layer ($A=10^{-11}$ J/m, $J_s=0.5$ T). The hard layer thickness as well as the soft layer thickness is 10 nm. For the calculation of the switching field distribution it is important to distinguish two cases.

1. **Uncorrelated layers:** In the first case we assume independent distributions of the anisotropy constant in the soft and the hard layer. The standard deviation of the anisotropy constant in the soft layer and hard layer is $\sigma_{soft} = 0.1 \times K_{soft}$ and $\sigma_{hard} = 0.1 \times K_{hard}$, respectively. Fig **4** shows the standard deviation of the switching field distribution ($\sigma_{\Delta H/H_s}$) and the mean value of the switching field ($\mu_0 H_s$) as function of the mean value of the soft layer anisotropy constant. Importantly, $\sigma_{\Delta H/H_s}$ shows a minimum for a mean a value of the anisotropy constant in the soft layer of $K_{soft} = 0.2$MJ/m³, which is 1/5 of the hard layer anisotropy. For practical application very helpful is the fact that also the switching field is minimized for anisotropy values of about 1/5 of the hard layer anisotropy. Hence, for the optimal value of the anisotropy constant in the soft layer (please consider the given value is the sum of the intrinsic anisotropy and the shape anisotropy) the standard deviation of the switching field distribution is decreased by almost a factor of ½. If the soft layer anisotropy is increased $\sigma_{\Delta H/H_s}$ increases until it reaches the maximum value if the soft layer anisotropy becomes equal to the hard layer anisotropy of $K_{soft} = 1.0$MJ/m³. In this case the structure describes a single phase media, where each grain is subdivided into two subgrains with independent anisotropy distributions. In this case the reduction of the switching field by a factor of $1/\sqrt{2}$ is in agreement with statistical arguments as explained in Ref [25],[26].
2. **Correlated layers:** The simulations of the previous paragraph were repeated by assuming correleated anisotropy distributions in the soft layer and the hard layer. Hence, if for example the anisotropy in the hard layer is 5% larger than the average anisotropy is the hard layer also the anisotropy in the soft layer is assumed to be larger by 5% compared to the average soft layer anisotropy. Fig 4 shows the results for the case of correlated anisotropy distribiutions in the hard and soft layer. If the average soft layer anisotropy becomes equal to the hard layer anisotropy (single phase media) the standard deviation of the switching field is almost 10% which agrees with the theoretical predictions for single phase media according to Eq. (5). Interestingly, again $\sigma_{\Delta H/H_s}$ is minimized if the average soft layer anisotropy is about 1/5 of the hard layer anisotropy.

In summary, independent if uncorrelated distributions in the soft layer and the hard layer or correlated distributions in the soft layer and hard layer are assumed, the standard deviation of the switching field distribution is smaller for bilayer media than in single phase media.

### *2.3. Precessional switching of bilayer media*

In the last section the magnetic properties such as the switching field and energy barrier, where investigated in a quasi static limit, where the external field is changed infinitely slowly. However, the field rise times in magnetic recording are in the order of several hundred picoseconds. These fast field rise times together with small damping constants in the media allow for precessional switching in composite media. The effect of precessional switching on switching times on exchange spring media was investigated in Ref [37]. It is shown that precessional switching is more pronounced in exchange spring media compared to single phase media. Reversal is reported for field pulses shorter than 25 ps [37]. Livshitz et al. investigated in detail the reduction of the switching field in composite media due to precessional switching[27]. It is found that precessional switching occurs in composite media for slower field rise times than in single phase media [27]. Furthermore, due to precessional switching the angular dependence of the switching field is significantly modified [27].

In Fig **6** phase diagrams are presented which show the regimes of switching and not-switching. The phase diagram is plotted as a function of the external field strength $\mu_0 H$ and the field pulse time $\tau$. Within a field rise time $t_r$ the field is increases linearly until the maximum field $\mu_0 H$ is reached. The maximum field is kept constant for a time $\tau$. After this time the field is decreased linearly to zero within the time $t_r$.

In Fig **6** the phase diagrams are shown for different angles θ, which denotes the angle between the external field and the film normal. Furthermore, the plots are done for different values of the field rise time $t_r$. The material properties of the investigated bilayer can be found in the figure caption of Fig **6**. Due to the small total thickness of the media not a full domain (180°) wall is formed but a 90° domain wall. The energy barrier of the investigated structure at zero field is for 48 $k_BT_{300}$ for a grain with circular basal plane with a diameter of 6 nm.

For fast field rise times of $t_r = 0.01$ ns and small angles θ = 10° it is interesting to note that the particle can be reversed with the shortest field pulses if the external field is just large enough to reverse the particle. These fastest switching modes for small external fields was also found in single phase media [28].

If for a field rise time of $t_r = 0.01$ ns the angle θ is increased to θ = 45° precessional switching is more pronounced. It leads to even faster reversal of the grain. For external fields between $\mu_0H = 2.0$T and $\mu_0H = 2.3$T the particle can even be switched for τ = 0. Hence, the particle is already reversed during the time when the field is increases from zero to $\mu_0H$ within the time $t_r$. Another interesting effect is that switching occurs much below the static switching field. The static switching field is $\mu_0H_{s,grain} = 1.37$T for θ = 45°. Due to the effect of precessional switching the particle can be reversed at fields of about $\mu_0H = 1.0$T.

If the field rise time is increased to $t_r = 0.1$ ns the phase diagrams in Fig **6** show that switching occurs within the time $t_r$. When the maximum field is reached a zero pulse length τ is sufficient to guarantee switching. Due to the larger field rise time for an angle θ = 45° no longer switching below the Stoner-Wohlfarth limit occurs. However, it is interesting to note that for an angle of θ = 10° precessional switching is still present. Even more important is that in the precessional switching regime the switching process can be faster for longer field rise times than for short field rise times [37]. For $t_r = 0.1$ ns, θ = 10° and for external fields between $\mu_0H = 1.6$ T and $\mu_0H = 2.6$ T the particle can even be switched for τ = 0, whereas the switching time in the same field interval for the fast field rise time of $t_r = 0.01$ is in the range of 0.3 ns to 0.5 ns.

For comparison phase diagrams of single phase media with a similar coercive field are shown in Fig **7**. It can be seen that the switching times are significantly longer than for the bilayer media. Furthermore, it is interesting to note that for the case of θ = 45° islands of switching and not switching can be found close to each other. Hence, more reproducible switching is expected for the bilayer case. The faster switching in the bilayer media is attributed to the larger anisotropy field in the hard layer. As a consequence the precessional frequency is larger in the bilayer media compared to the single phase media.

# 3. Graded Media

## *3.1. Optimization of graded Media*

The concept of exchange spring media can be further improved if the number of layers is increased. In a multilayer structure with continuously increasing anisotropy from layer to layer it was shown theoretically that the switching field can be decreased to an arbitrarily small value while keeping the energy barrier (thermal stability) constant [29]. This structure will be called in the following graded media.

The basic principle of graded media is illustrated in Fig **8**. For a particle in a potential well, the maximum force that is required to move it from one minimum to another minimum depends on the gradient. The gradient can be decreased by scaling the energy landscape in the horizontal direction. The thermal stability which is determined by the energy barrier is not influenced by the scaling. In a magnetic system the *y*-axis in the plot is the energy for different magnetic configurations. The *x*-axis can be regarded as the component of the magnetic moment $m_z$ parallel to the easy axis ($m_z = V*M_z$), where V is the volume of the particle and $M_z$ the magnetization. For a single domain particle $m_z = V * M_z = V * M_s * \cos(\theta)$, where θ is the angle between the easy axis and the magnetization. For incoherent switching the $m_z$ can be regarded as the position of the centre of the domain wall [30]. In the magnetic picture the scaling of the energy landscape can be realized by the introduction of magnetic layers with different magnetic anisotropy constants. With increasing total layer thickness the total magnetic moment increases. In turn the maximum slope of the energy landscape decreases. Therefore, the field (in the mechanical analog the force) that is required to switch the particle can be decreased without changing the energy barrier. Also in small magnetic particle that reverse via homogenous rotation the coercive field can be decreased while keeping the energy barrier constant if the total magnetic moment is increased by increasing the particle length and reducing at the same time the anisotropy constant *K*. However, this only works until to a critical particle length of $t_{crit} = 4\sqrt{A/K}$. Above this critical length the particle reverses via the formation of a domain wall resulting in an energy barrier and coercive field that is independent on the particle length. For FePt alloys for example the critical thickness is reached for $t_{crit} \approx 4.8 nm$.

The switching field in graded media can be minimized if the anisotropy constant is scaled quadratically as a function of the distance from the top of the layer to the bottom, $K(z) = \frac{K_{max}}{t_g^2} z^2$. The maximum anisotropy at the bottom of the layer is $K_{max}$. The total layer thickness is $t_g$. For a quadratic anisotropy profile the energy increases linearly as function of $m_z$ until is sharply decreases when the entire grain is reversed [29,30].

The switching field for a graded media is given by [29],

$$\mu_0 H_{s,grain} = (2/J_s) \times \sqrt{AK_{max}}/t_g, \qquad (6)$$

The required write field can be reduced by increasing the spontaneous polarization. However, care has to be taken because large values of the magnetization produces larger demagnetizing field lowering the thermal stability. The thermal stability of one grain of the media can be estimated by the energy barrier

$$\Delta E = \Delta E_0 (1 - H/H_{s,grain})^n \qquad (7)$$

Here, $\Delta E_0$ is the energy that is required to switch the grain from up to down at zero field. $H$ is the field that acts on the grain and $H_{s,grain}$ is the switching field of just one grain without interactions. The most unstable grain is the grain in the centre of the bit, where the largest demagnetizing field occurs. For the field acting on the centre grain follows,

$$\mu_0 H = N_{eff} J_s - \mu_0 H_{ex} \qquad (8)$$

It can be seen that increasing the magnetic polarization does not only decrease the switching field of one grain, but it also decreases the thermal stability of the grain due to larger self demagnetizing fields. Although the demagnetizing field is partly compensated by the exchange field it poses important limits to the maximum values of the magnetic polarization. The optimal value for the magnetization in order to obtain the maximum energy barrier of a grain for a constrained saturation field follows from an optimization problem [1]. Charap et al. calculated the thermal stability for different materials with different values of $J_s$ for longitudinal recording [4]. Analysis of the optimal choice of $J_s$ in perpendicular recording can be found in Ref [5,6].

For graded media the optimization problem can be solved more easily since the exponent $n$ in Eq. (7) can be assumed to be very close to one [23].

In order to solve the optimization problem for graded media we express the energy barrier at zero field as a function of the switching field of one grain

$$\Delta E_0 = 4F\sqrt{AK} = 2FJ_s H_{s,grain} t_g, \qquad (9)$$

where $F$ is the cross-sectional area of one grain and $t_g$ the total thickness of the graded media. Combing (9) and (8) we get

$$\Delta E = 2FJ_s H_{s,grain} t_g (1 - \frac{N_{eff} J_s - \mu_0 H_{ex}}{\mu_0 H_{s,grain}}) \qquad (10)$$

In order to fulfill the constraint of a constant saturation field we express the switching field of one grain ($H_{s,grain}$) as a function of the saturation field $\mu_0 H_{sat}$ using Eq. (1). Writeability of the medium can be guaranteed if the saturation field is equal to the maximum head field. $\mu_0 H_{sat} := \mu_0 H_{head}$. In the following, we obtain the energy barrier of the grain in the centre of a bit as a function of the maximum head field $H_{head}$, the exchange field $H_{ex}$ the total layer thickness, the intragrain exchange constant $A$, and the magnetic polarization $J_s$. In Fig **9** the energy barrier of different graded media with different values of the magnetic polarization are plotted. All structures have the same saturation field and hence the same writeability. It is shown that for the given material parameters the optimum structure has a magnetic polarization of $J_s = 0.8T$.

The optimum value of the magnetic polarization can also be obtained analytically by setting the first derivative of the energy barrier (Eq. (10)) with respect to $J_s$ to zero. We obtain for optimum value of the magnetic polarization

$$J_s = \frac{1}{4N_{eff}} (2\mu_0 H_{ex} + \mu_0 H_{head}) \qquad (11)$$

After obtaining the optimum magnetic polarization Eq. (10) and Eq. (1) can be used to calculate the energy barrier.

### *3.2. Gain of graded Media compared to single phase media*

In order to compare the benefit of graded media compared to single phase media different figure of merits where proposed. Victora et al proposed as the figure of merit the factor $\xi$

which is the ratio of the energy barrier over the switching field as given by

$$\xi = \frac{2\Delta E}{H_{s,grain} J_s V},$$

where $V$ is the volume of the magnetic grain.

For single phase media, where the grains reverse homogeneously the ratio is one. Since the ratio is independent of the thickness of the sample for single phase media it can be seen that this figure of merit has its limitation. A single phase medium with a thickness of 6 nm is definitely favorable due to a higher stability than a single phase media with a thickness of 3 nm. However, both structures show the same figure of merit.

In order to take care of this concern we define the figure of merit the ratio of the energy barrier of the graded media over the energy barrier of the single phase media with the constraints that the single phase media has the same magnetization, the same thickness and the same switching field as the graded structure.

Fig **10** shows the energy barrier for the graded media and the single phase media as a function of the total layer thickness. For thin single phase media the energy barrier increases as the layer thickness increases. In this region the lowest energy barrier follows from the Stoner-Wohlfarth theory as $\Delta E_{SW} = KV = KFt_s$, where $t_s$ is the layer thickness of the single phase media. For thick layers the lowest energy barrier is no longer given by homogeneous rotation but it is favourable that the particle reverses via the formation of a domain wall. The corresponding enegry barrier is $\Delta E_{domain} = 4F\sqrt{AK}$, which is independent on the layer thickness. The critical thickness which defines the transtion between homogebous rotation and the formation of a domain wall occures if $\Delta E_{SW} = \Delta E_{domain}$. Hence, for a thickness of $l_c = 4\sqrt{A/K}$ both energy barriers become equal. For single phase media thicker than $l>l_c$ the energy barrier saturated and the thermal stability can not further improved by increaseing the film thickness.

The situation is different for graded media. Here, for all layer thicknesses the energy barrier increases linearly as a function of the layer thickness as given by Eq. (9).

As a consequence for layer thicknesses $l<l_c$ the energy barrier of the graded media is larger by a factor of 4 compared to the single phase media [32]. However, care has to be taken that for too thin graded media reversal does no longer happen with the formation of a domain wall and Eq. (9) is no longer valid. For $l>l_c$ the energy barrier of graded media continously increases while it stays constant for the single phase media. Hence, in principle the gain of graded media goes to infinity for $l \rightarrow \infty$.

For practical application the maximum layer thickness is bounded by several factors, such as the decay of the head field for larger layer thickness, the requirement of a columnar structure which is hard to maintain for very thick layers, and the high coercive materials which would be required for very thick graded media.

### *3.3. Switching field distribution*

In order to investigate the switching field distribution in magnetic multilayers and graded media a clear understanding of the underlying distribution of the anisotropy constant in each layer is required. In particular it is important to distinguish two cases:

1. **Uncorreleated layers:** The investigations of Section 2.2 are repeated for the graded media with a total thickness of 20 nm. For 400 graded media grains the switching fields are calculated. One grain of the graded media is constructed with 20 layers. The average anisotropy constant of a grain increases quadratically from layer to layer. The maximum anisotropy is $K_{1,\mathrm{max}} = 2.0 \mathrm{MJ/m^3}$. In each layer a distribution of the anisotropy of 10% is assumed. During reversal a domain wall is formed. The domain wall is extended over several layers. Hence, the domain wall is exposed to an anisotropy which is the average over several layers. As a consequence the standard deviation of the switching field distribution is drastically reduced to be only 3% of the average switching field. Since, it is not clear if completely uncorrelated distributions of the anisotropies in each layer are realistic the switching field distribution is also investigated for the case of correlated layer.
2. **Correlated layers:** Here, we assume a perfect quadratic increase of the anisotropy constant from layer to layer. A distribution of the anisotropy constant in one grain is realized by scaling the anisotropy constant in each layer with the same factor. In order to calculate the switching field distribution we use Eq. (6) and assume a variation of the hard layer anisotropy by $\Delta K_1$. In contrast to single phase media the switching field of graded media depends on the square root of the hard layer anisotropy. As a consequence the change of the switching field due to a variation of the anisotropy constant $K_1$ is given by

$$\frac{\Delta H}{H} = \frac{\mu_0 H_{s,grain,graded}(K_1 + \Delta K_1) - \mu_0 H_{s,grain,graded}(K_1)}{\mu_0 H_{s,grain,graded}(K_1)} = \frac{1}{2}\frac{\Delta K_1}{K_1} \quad (12)$$

Comparing Eq. (12) and Eq. (5) shows that the switching field distribution of graded media is smaller by a factor of two compared to the single phase media where the switching field depends linearly on the anisotropy constant.

### *3.4. 5 - 10 Tbit/inch² design:*

In this paragraph present a graded media design which should be capable for ultra high density recording up to 5-10 Tbit/inch². The recording layer is a graded media, where the anisotropy increases from zero to $K_{max} = 6.6 MJ/m^3$ according to $K(z) \propto z^{1.5}$. The profile of the anisotropy takes into account for the decay of the head field with increasing distance $z$ from the top of the media The total layer thickness is $t$ = 20 nm and the spontaneous polarization is $J_s$ = 0.8 T. In order to guarantee fast switching of the grains a damping constant of $\alpha$ = 0.1 is assumed. The energy barrier of one grain in the centre of the bit was calculated using the nudged elastic band method [16]. The obtained energy barrier is $\Delta E = 60 k_B T_{300}$. The write field requirements to reverse the grains is that the maximum head field at the top of the media is 1.8 T . We have assumed that it decays to 1.0 T at the bottom of the media (20 nm). The average grain diameter of the structure is $d$ = 3.2 nm, which is sufficient small to allow for recording at ultra areal densities up to 5 - 10 Tbit/inch².

## 4. Challenges of exchange spring and graded media

In the following section we will point out the problems and the challenges which have to be overcome to realize media that are capable to store information with areal densities of several Tbit/inch². The fabrication of exchange spring media demands for granular media with superior properties. In particular the following demanding properties are required:

- **Small grains sizes:** The concept of exchange spring and graded media relies on thermally stable grains with small grain diameters. The fabrication of granular media with grain diameters of the order of 4 nm and smaller is technologically very challenging.
- **Hard $K_1$ materials are required:** To make fully use of the concept of graded media materials with $K_1$ values close to that of FePt are required. If we assume that the available writefield is about 1.5T one can switch a bilayer media and a trilayer media with a maximum anisotropy in the hardest layer of about $K_1 = 1.4$ MJ/m³ and $K_1 = 2.8$ MJ/m³, respectively. Hence, already for the trilayer media anisotropies beyond these of Co and CoPtCr alloys have to be used, which show anisotropies of about $K_1 = 0.2$ - 0.45 MJ/m³ [2].

The reversal process in exchange spring media differs significantly from single phase media. As a consequence as already pointed out **precessional switching** occurs for relative slow rise times of the order of 0.1 ns. Precessional switching helps to decrease the switching field and switching as fast as 20 ps can be realized. However, precessional switching also changes the angular dependence of the switching field [27]. Due to precessional switching the minimum switching field occurs for field angles between the easy axis and the external field of about θ = 45°. This is different to most exchange spring media in the case of strong damping which show in the ideal case a angular dependence according to the Kondorski relation (l/cos(θ) law) [33]. A Kondorski relation of the angular dependence of the pinning field is argued to improve the problem of adjacent track erasure [5,6]. Hence, it is possible that precessional switching has to be avoided in order to decrease the problem of adjacent track erasure.

## 5. Open Questions

Although composite media and exchange spring media are investigated by varies groups and excellent papers are published a large number of open and interesting questions still remain. In the following I will mention only two of them.
As a drawback of graded media the relative large larger thickness of 10 nm to 20 nm is under debate. As an argument it is mentioned that high density recording requires superior field gradients, which only can be realized very close to the pole tip. As a consequence media designs with thicknesses of 10 nm or smaller are proposed. An interesting question is if in graded media due to the different reversal mode compared to single phase media larger layer thicknesses up to 20 nm can be tolerated. In graded media a nucleation is formed at the top of the columnar grains if the head field exceeds the switching field. A grain can only be reversed if a nucleation at the top is initially formed. Hence, the regions where nucleations can be formed define which grains are going to be reversed. In conventional perpendicular recording media sharp bit transitions require a large head field gradient over the entire thickness of the media. Due to the domain wall reversal in graded media a one can speculate that high field gradients are only required close to the pole tip. The region where a high field gradient is required is determined by the nucleation volume. Detailed studies to explore the benefit of the head field gradient are an interesting topic for future research.

Composite media design where proposed with negative anisotropy instead of the soft layer [34]. As a consequence the magnetization is oriented in plane in the negative $K_u$ material. As pointed out in Fig **2** the switching field can not be further decreased by decreasing the anisotropy of the soft layer to negative values. Furthermore the thermal stability is slightly decreased with negative $K_u$ values. However, if in combination with the negative $K_u$ material, a material with large magnetization is used a further reduction of the switching field can be obtained. Due to the in plane magnetization the large magnetization does not lead to high demagnetizing fields in the layer with the in plane anisotropy. The large demagnetizing field in perpendicular recording limits the saturation magnetization to values around 0.5 T – 0.7 T.

## 6. Summary

The potential of graded media and exchange spring is discusses and compared with single phase media. In bilayer media the influence of the soft layer anisotropy on the ratio energy barrier over coercive field is investigated. It is shown that an inplane anisotropy in one of the layer is not beneficial in order to increase the ratio energy barrier over coercive field. Micromagnetic studies were performed in order to show the influence of anisotropy distributions in the soft and hard layer of exchange spring media on the coercive field. The two extreme situations of (i) completely correlated and (ii) completely uncorrelated anisotropy distributions in the soft layer and the hard layer are investigated. In both situations it was found that the switching field distribution in exchange spring bilayers can be significantly reduced. In the case of correlated distributions the standard deviation of the switching field distribution of a magnetic bilayer is reduced by about a factor of 1.6 compared to a single phase media. In the case of uncorrelated distribution a fair comparison requires the calculation of the switching field distribution of a single phase media which is subdivided into two subgrains. A single phase media with two subgrains with independent distributions of the anisotropy constants has a switching field distribution that is reduced by a factor of $1/\sqrt{2}$ compared to a single phase media with only one $K_1$ value per grain. The simulations on the bilayer media, assuming uncorrelated anisotropy value distributions show a reduction of the switching field distribution by a factor of ½ compared to a single phase media. Hence, bilayer media have a switching field distribution that is smaller by a factor of about $1/\sqrt{2}$ compared to the single phase media with two subgrains.

The switching speed of bilayer media is investigated by applying field pulses with different strength, pulse length and field rise time. It is shown that precessional switching in magnetic bilayers allows switching well below the static switching field. For a field risetime of 0.01 ns and a field that is applied at an angle of 45° with respect to the easy axis the particle can be switched for a total length of the field pulse of 10 ps. This is significantly faster than for a single phase media with comparable coercive field.

The concept of graded media where the anisotropy constant varies from layer to layer quadratically is reviewed. An analytical model is presented that allows for calculating the optimal value of magnetic polarization in order to maximize the ratio energy barrier over coercive field. Due to the fact that the energy barrier decays linearly as a function of the external field strength, $\Delta E = \Delta E_0 \left(1 - H/H_c\right)$ the thermal stability of grains of graded media are insensitive to magnetic fields, such as fields of permanent magnets located close the hard disc, or internal demagnetizing fields. As a consequence magnetic materials with slightly higher magnetization can be used since the self demagnetizing field leads only to a moderate reduction of the thermal stability of the grains in the centre of a bit. The optimum value of the magnetic polarization which depends on the intergranular exchange can be obtained by using Eq. (11). For reasonable material parameters the optimal ratio energy barrier of coercive field is obtained for $J_s = 0.8$ T.

The gain of graded media compared to single phase media is reviewed. If a graded media is compared with a single phase media of same size and same magnetic polarization, the gain is a factor of 4 provided the single phase media is sufficient small to reverse via homogenous rotation. For future recording media with high $K_1$ values single phase media exceeding a thickness of 5 nm will reverse via the formation of a nucleation. In this case the gain in thermal stability of graded media compared to single phase media scales linearly with the layer thickness. In all investigated situations the switching field was calculated in the quasi static limit. Hence, a reduction of the switching field due to precessional switching was not taken into account.

The switching field distribution of graded media is reduced due to the fact that the switching field depends on the square root of the anisotropy constant in the hardest layer. If completely correlated distributions between the layers of a graded media are assumed the standard deviation of the switching field is reduced by a factor of two compared to a single phase media.

The financial support of the Austrian Science Fund P19350 and P20306 is acknowledged.

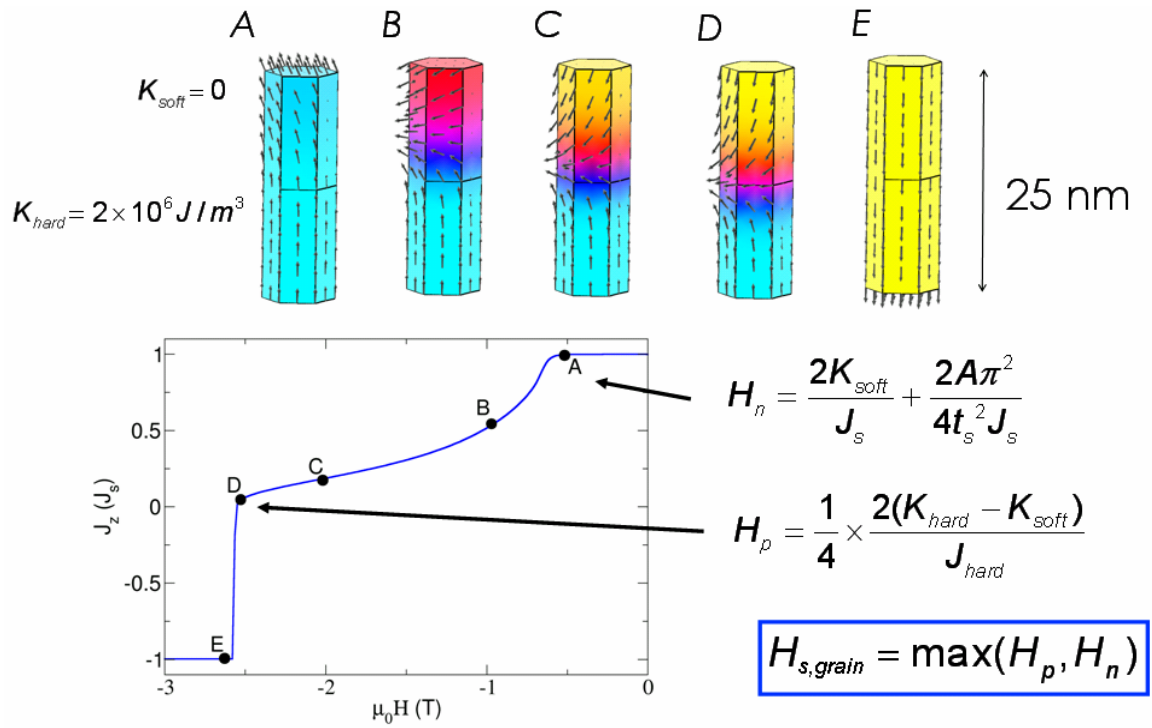

Fig 1: Compression of a domain wall at the soft/hard interface after it was nucleated in the soft end. The switching field of one grain ($H_{s,\text{grain}}$) is determined by the maximum of the nucleation field $H_n$ and the pinning field $H_p$.

$$H_n = \frac{2K_{soft}}{J_s} + \frac{2A\pi^2}{4t_s^2 J_s}$$

$$H_p = \frac{1}{4} \times \frac{2(K_{hard} - K_{soft})}{J_{hard}}$$

$$H_{s,\text{grain}} = \max(H_p, H_n)$$

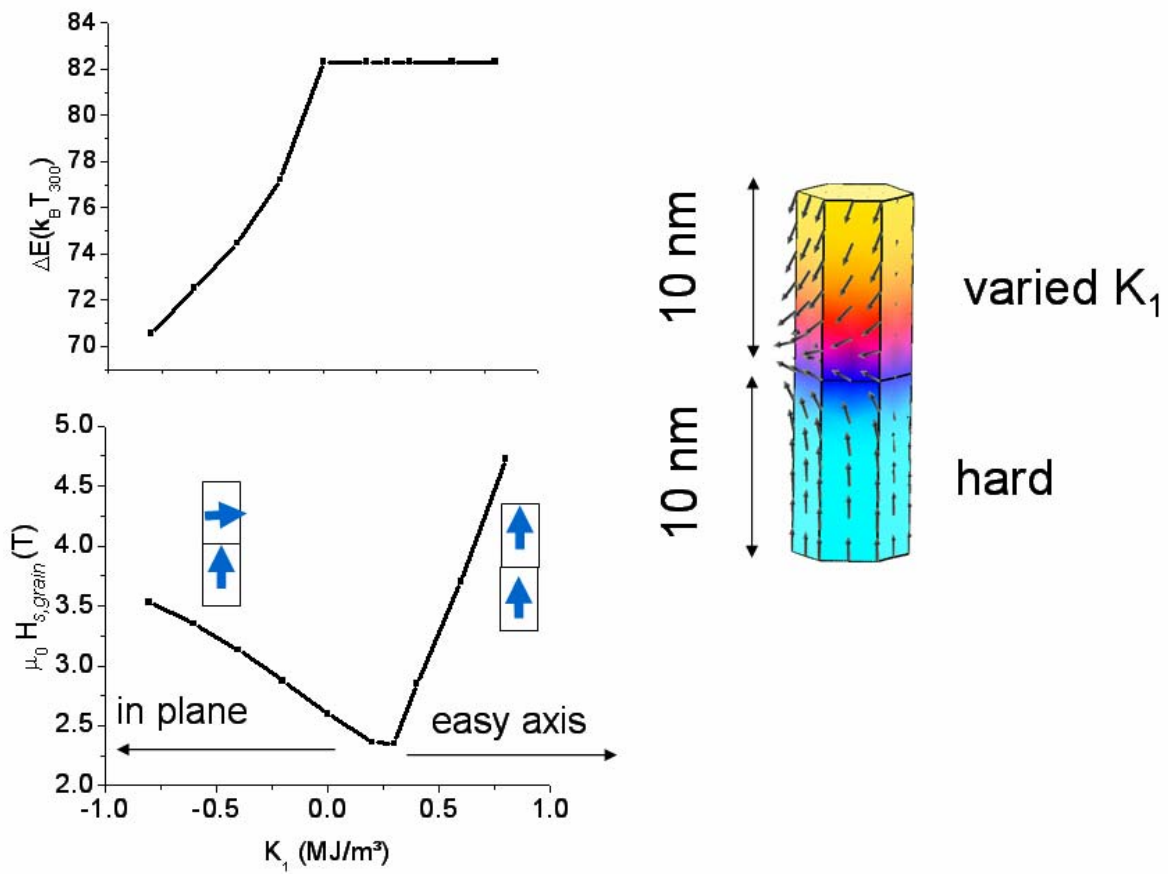

Fig 2: Influence of the strength of the top grain anisotropy on the switching field and the energy barrier of a bilayer structure with a grain diameter of 5 nm. The magnetic properties in the hard and soft grains are: A=$10^{-11}$ J/m, $J_s$ = 0.5 T. The hard layer anisotropy is $K_{hard}$ = $2 \times 10^6$ J/m³. The soft layer anisotropy is varied from a positive value to a large negative value which leads to an inplane anisotropy.

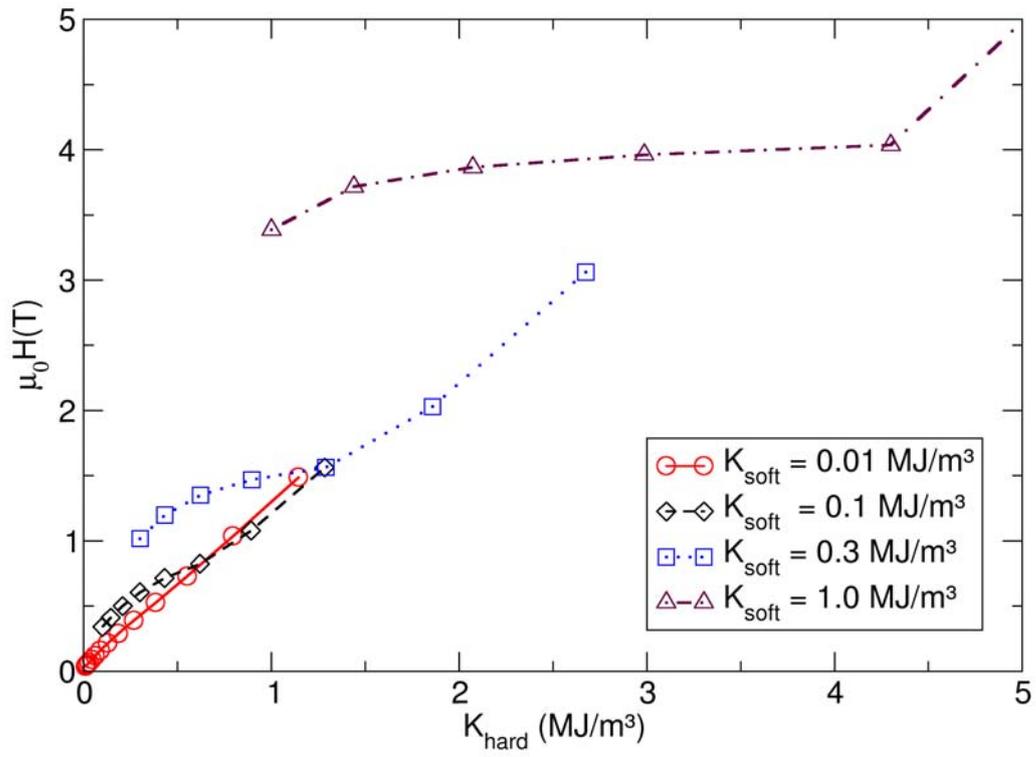

Fig 3: Switching field of one grain of a bilayer media as function of the hard layer anisotropy constant ($K_{hard}$). The simulations are repeated for different soft layer anisotropies ($K_{soft}$).

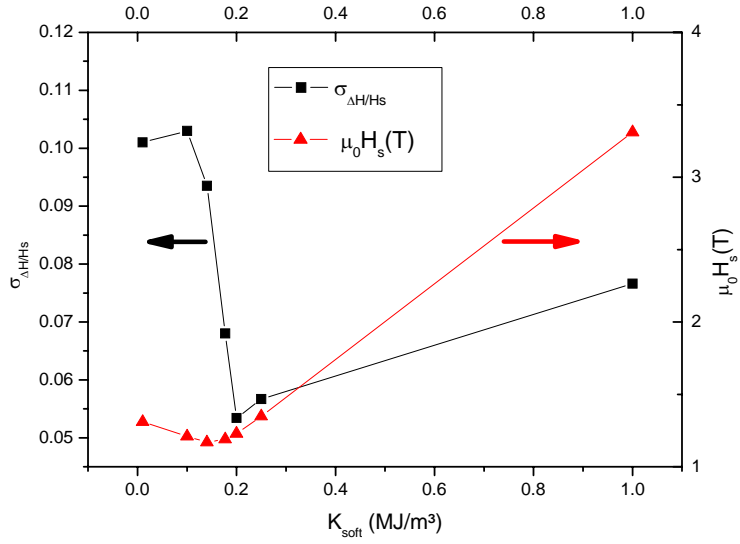

Fig 4: Dependence of the standard deviation of the switching field distribution ($\sigma_{\Delta H / H_s}$) and the mean value of the switching field ($H_s$) of a magnetic bilayer as function of the mean value of the soft layer anisotropy. In the soft layer and the hard layer a normal distribution of the anisotropy constant was assumed. The standard deviation of the normal distribution of the anisotropy in the soft and hard layer is 10% of the mean value. The distribution of the anisotropy constant in the soft layer and hard layer are completely uncorrelated.

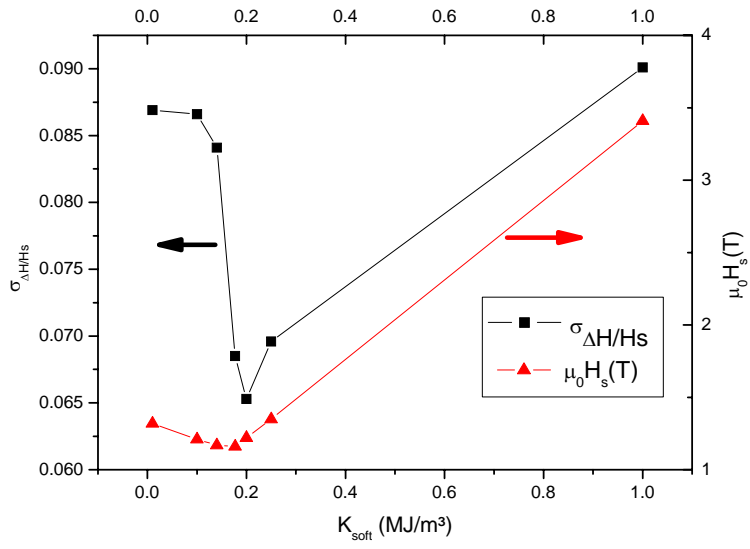

Fig 5: Same as Fig 4 except that the distributions of the soft layer anisotropy constants and the hard layer anisotropy constants are completely correlated.

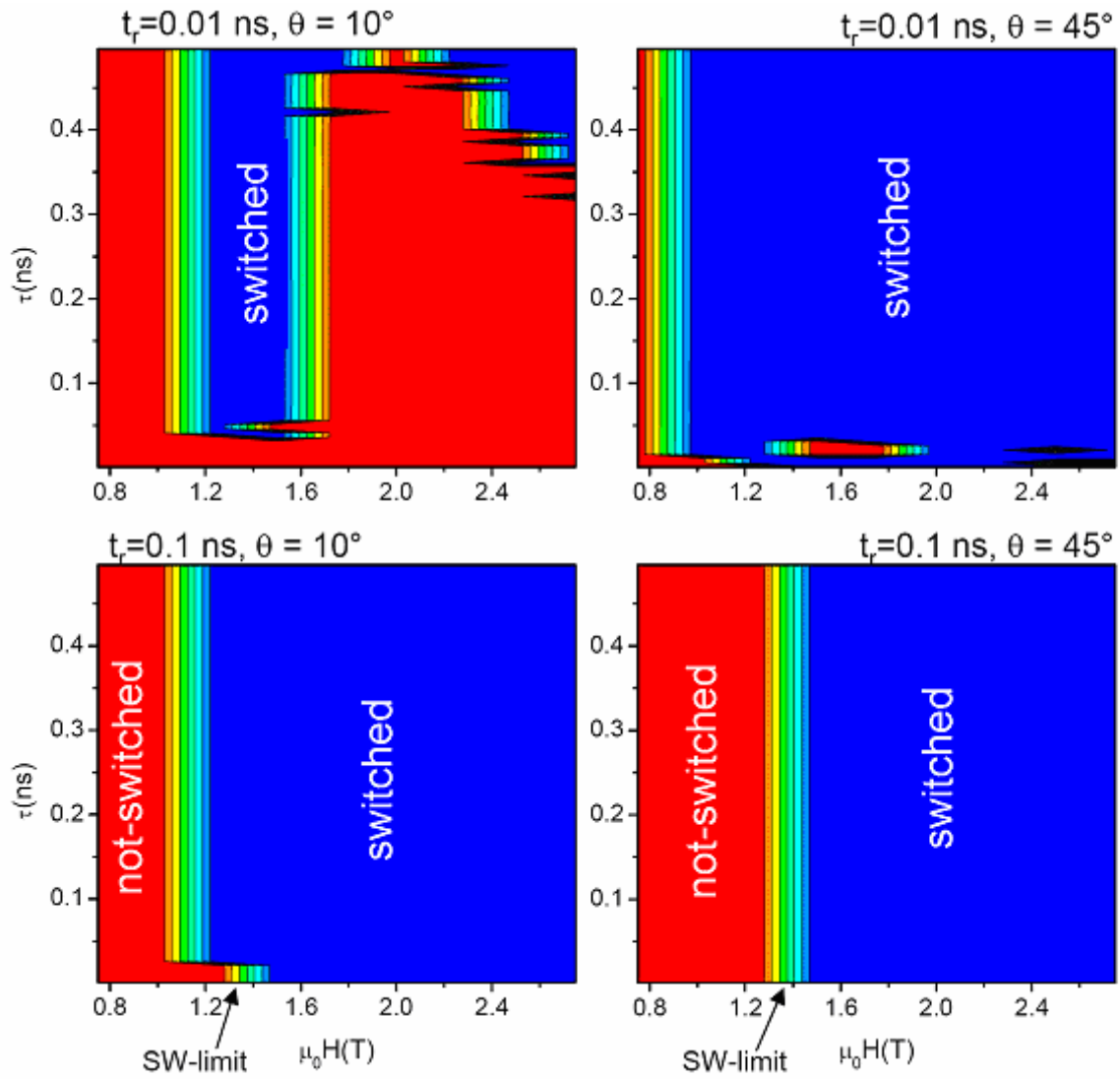

Fig 6: Phase diagram of the switching event for a bilayer media with a total thickness of 11 nm. The soft layer ( 5.5 nm) is assumed to be perfectly soft ($K_1 = 0$, $A=10^{-11}$ J/m, $J_s = 0.5$ T). The anisotropy of the hard layer is $K_1 = 1$ MJ/m³. The field rise time $t_r$ (in nanoseconds) and the angle θ between the easy axis an the film normal is varied. The damping constant $\alpha = 0.02$.

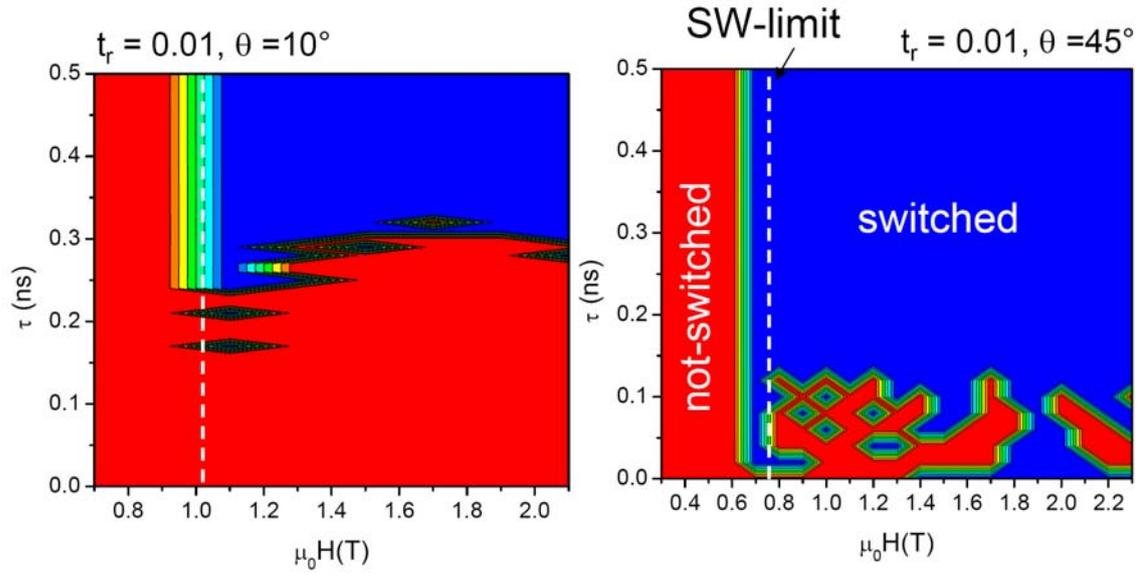

Fig 7: For comparison a phase diagram of the switching event for a single phase media with a total thickness of 11 nm is shown. The anisotropy of the hard layer is $K_1 = 0.3$ MJ/m³. The field rise time $t_r = 0.01$ ns. The angle θ between the easy axis an the film normal is varied. The damping constant $\alpha = 0.02$.

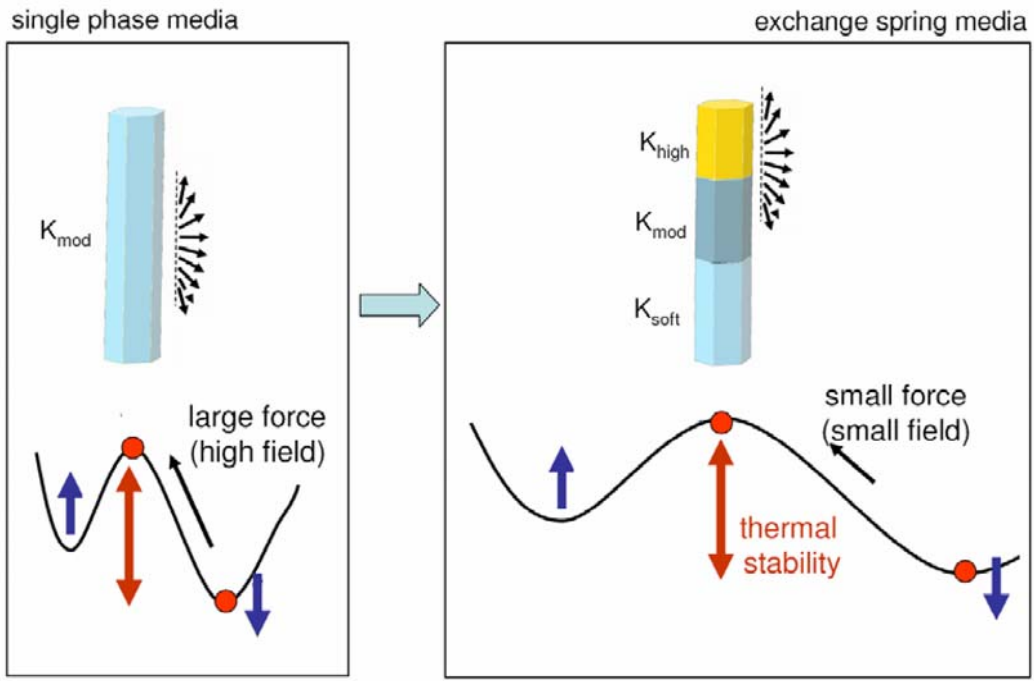

Fig 8: Basic principle of exchange spring media for magnetic recording. Similar to a mechanical analogy it is possible to keep the same thermal stability (energy barrier) between two stable states but changing the force (field) that is required to change the state of the system.

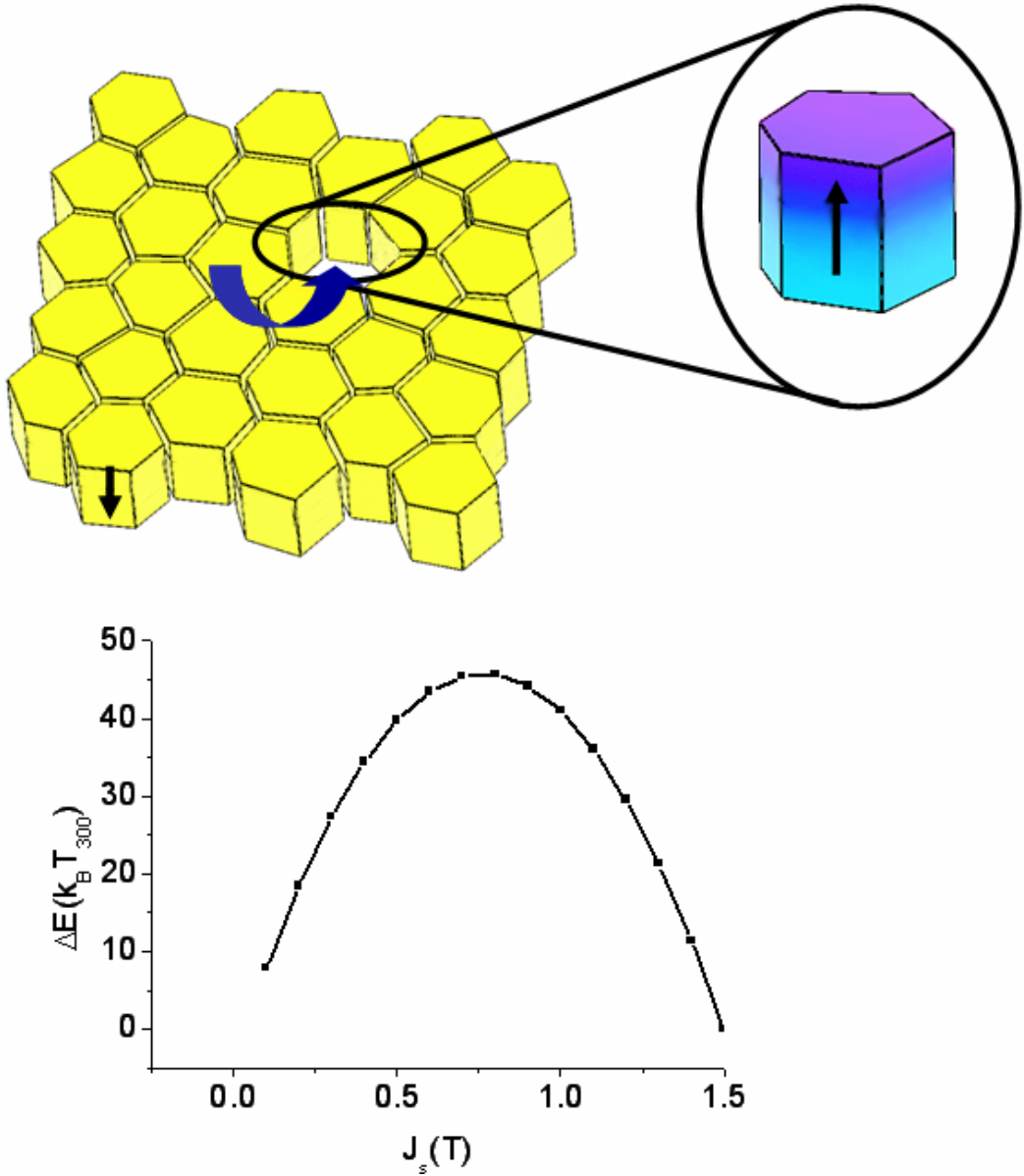

Fig 9: (top) Illustration of the mean field approach. The neighboring grains of the grain of interest $g_i$ produce a magnetic field in the region of $g_i$. In the following it is assumed that the field in the hole of the region $g_i$ is $J_s N_{eff}$, whereas the demagnetizing factor of the film is assumed to be $N_{eff} = 0.8$.
(bottom) Results of an analytical mean field model to predict the optimum value of the magnetic polarization in the graded media in order to maximize the energy barrier. The saturation field is constrained to be $\mu_0 H_{sat} = 1.4$ T. The intergrain exchange field between the grains is $\mu_0 H_{ex} = 0.5$ T, the intragrain exchange constant $A = 10^{-11}$ J/m, the demagnetizing factor of the film is $N_{eff} = 0.8$, the film thickness is 20 nm and the grain diameter is 3 nm. For a constrained saturation field the maximum thermal stability is obtained for $J_s = 0.8$ T.

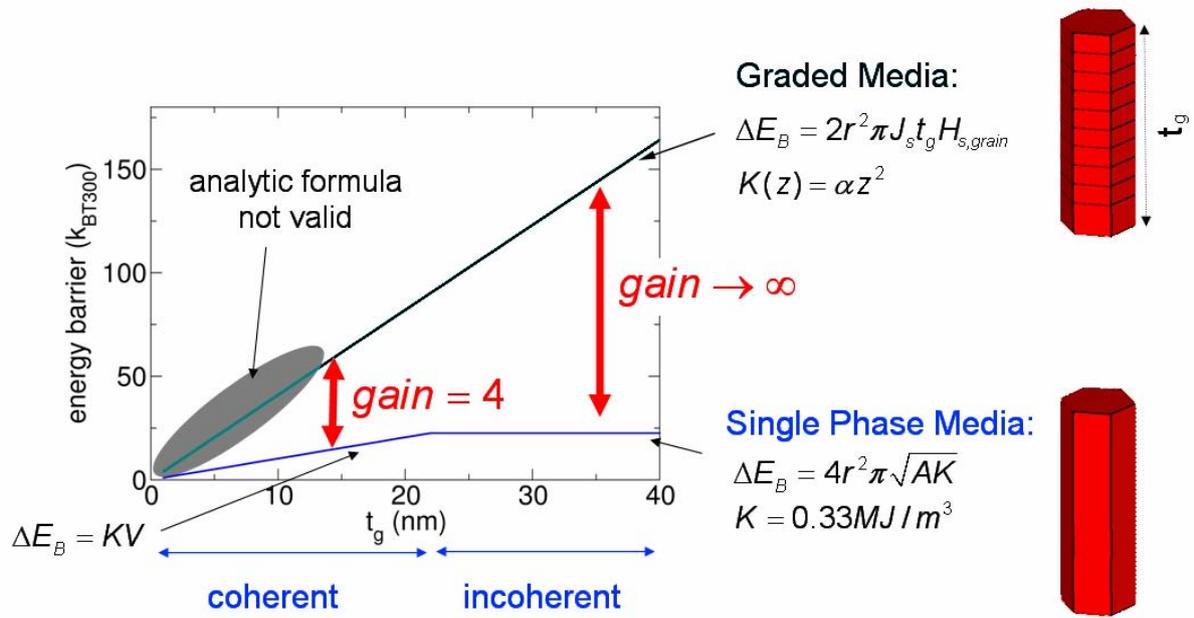

Fig 10: Comparison of the energy barrier (thermal stability) of a single layer media and a graded media. Both grains have grain diameter of 4 nm and the same magnetic polarization of $J_s = 0.5$ T. The switching field of both structures is $\mu_0 H_{s,grain} = 1.7$ T

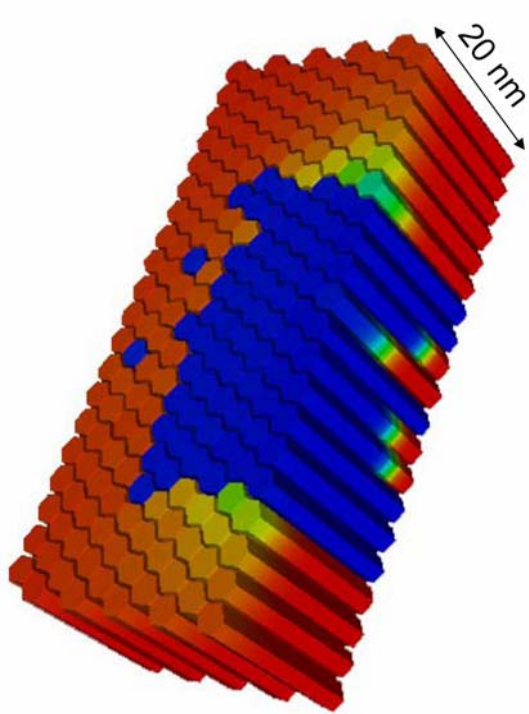

| Tapered head / Planer head | |
|---|---|
| top of media | $\mu_0 H_p = 1.8$ T |
| bottom of media | $\mu_0 H_p = 1.0$ T |

| Multilayer media | |
|---|---|
| grain size | **3.2 nm** |
| thickness | 20 nm |
| $J_s$ | 0.8 T |
| $K(z)$ | $\alpha\ z^{1.5}$ |
| anisotropy | 0.1 MJ/m³ to 6.6 MJ/m³ |
| Energy barrier | 60 $k_B T$ |

Fig 11: Graded media design which has the potential to support areal densities up to 10 Tbit/inch². The left images shows transient states during the write process.

# References

[1] James M. Chirico, „Seagate Outlines the Future of Storage", URL: http://www.hardwarezone.com/articles/view.php?cid=1&id=1805&pg=1

[2] D. Weller, A. Moser, L. Folks et al, IEEE Trans. Magn., 36 (2000) 10.

[3] J. Miles, D. Mckirdy, and R.Wood, IEEE Trans. Magn., 39 (2003) 1876–1890.

[4] K.Z. Gao, H.N. Bertram, IEEE Trans. Magn., 39 (2003) 704.

[5] R. H. Victora and X. Shen, IEEE Trans. Magn., 41 (2005) 2828-2833.

[6] D. Suess, T. Schrefl, S. Fahler, et al, Appl. Phys. Lett., 87 (2005) 012504.

[7] A. Dobin and H. J. Richter, 89 (2006) 062512.

[8] F. Garcia-Sanchez, O. Chubykalo-Fesenko, O. Mryasov, R. W. Chantrell, and K. Yu. Guslienko, Appl. Phys. Lett. 87 (2005) 122501.

[9] J. P. Wang, W. K. Shen, and J. M. Bai, IEEE Trans. Magn. 41 (2005) 3181-3186.

[10] N. F. Supper, D. T. Margulies, A. Moser et al. J. Appl. Phys., 99 (2006) 08S310.

[11] H. Kronmüller and H. R. Hilzinger, J. Magn. Magn. Mater., 154 (1976) 3-10.

[12] K. Y. Guslienko, O. Chubykalo-Fesenko, O. Mryasov, R. Chantrell, and D. Weller, Phys. Rev. B, 70, (2004) 104405.

[13] F. B. Hagedorn, J. Appl. Phys., 41 (1970) 2491-2502.

[14] P. Loxley and R. L. Stamps, IEEE Trans. Magn. 37 (2001) 1998 – 2100.

[15] H. Kronmuller and D. Goll, Physica B, 319 (2002) 122-126.

[16] R. Dittrich, T. Schrefl, D. Suess, et al. J. Magn. Magn. Mater. 250 (2002) 12-19.

[17] J. Dean, M. A. Bashir, A. Goncharov, et al., Appl. Phys. Lett. 92, (2008) 142505

[18] J.W. Harrell, IEEE Trans. Magn., 37, (2001) 533.

[19] R. Victora, Phys. Rev. Lett. 63, (1989) 457–460.

[20] H. Pfeiffer, Phys. Stat. Solidi(a) 118, (1990) 295–306.

[21] D. Suess, S. Eder, J. Lee, et al., Phys. Rev. B 75, 174430 (2007).

[22] D. Goll, S. Macke, and H. N. Bertram, Appl. Phys. Lett, 90, (2007) 172506.

[23] D. Suess, G. Zimanyi, T. Schrefl et al, Appl. Phys. Lett (2008).

[24] A. Dobin, Intermag Conference, Madrid (EF-01) 2008.

[25] Y. Sonobe, K. K. Tham, T. Umezawa, C. Takasu, J. A. Dumaya, and P.Y. Leo, J. Magn. Magn. Mat. 303 (2006) 292–295.

[26] D. Suess, J. Fidler, H. S. Jung, E. M. T. Velu, W. Jiang, S. S. Malhotra, G. Bertero, T. Schrefl, IEEE Trans. Magn. (2008), submitted.

[27] B. Livshitz, A. Inomata, H. N. Bertram, and V. Lomakin, Appl. Phys. Lett. 91 (2007) 182502.

[28] D. Suess, T. Schrefl, J. Fidler, IEEE Trans. Magn. 37 (2001) 1664.

[29] D. Suess, Appl. Phys. Lett., 89 (2006) 113105.

[30] Z.H. Lu, P. Visscher, W. H. Butler, IEEE Trans. Magn., 43 (2007) 2941 – 2943.

[33] E. Kondorski, Phys. Z. Sowjetunion 11, (1937) 597.

[34] Advanced Granular-Type Perpendicular Recording Media — New materials and new stacked layer form —M. Takahashi, and S. Saito, PMRC, Tokio (2007)